\documentclass[11pt]{article}
\usepackage{graphicx} \usepackage{amsmath} \usepackage{amsfonts}
\usepackage{amssymb}

\def\dsp{\displaystyle}
\def\d{\mbox{\rm d}}

\def\dt{\mbox{\rm d}t}

\def\f{\frac}

\def\half{\mbox{$\frac{1}{2}$}}

\def\p{\dsp\partial}

\def\dddot#1{\mathinner{\buildrel\vbox{\kern5pt\hbox{...}}\over{#1}}}
\def\dm#1#2{\f{\dsp \d #1}{\dsp \d #2}}
\def\pa#1#2{\f{\dsp \p#1}{\dsp \p#2}}

\def\etal{{\it et al }}
\def\ie{{\it ie }}

\def\re#1{(\ref{#1})}
\def\viz{{\it videlicet }}

\begin{document}


\thispagestyle{empty}

\title{Jacobi's last multiplier and the complete symmetry group of
the Euler-Poinsot system}

\label{firstpage}

\author{MC Nucci~$^ {\dag} $ and PGL\ Leach~$^{\dag\,\ddag}$}

\date{$^\dag$ Dipartimento di Matematica e Informatica\\
~~Universit\`a di Perugia, Perugia 06123, Italy\\[2mm]
$^\ddag$ permanent address: School of Mathematical and Statistical
Sciences\\~~University of Natal, Durban, South Africa 4041}

\maketitle

\begin{abstract}
\noindent The symmetry approach to the determination of Jacobi's
last multiplier is inverted to provide a source of additional
symmetries for the Euler-Poinsot system.  These additional
symmetries are nonlocal.  They provide the symmetries for the
representation of the complete symmetry group of the system.
\end{abstract}

\section{Jacobi's Last Multiplier}

The method of Jacobi's last multiplier \cite {Jacobi 44 a, Jacobi
44 b, Jacobi 45} (see also \cite {Koenigsberger 04} [p 320, p 335
and pp 342-347] for a summary of these three papers of Jacobi)
provides a means to determine an integrating factor, $M $, of the
partial differential equation
\begin {equation}
Af= \sum_{i = 1} ^n a_i\pa {f} {x_i} = 0 \label {1.1}
\end {equation}
or its equivalent associated Lagrange's system
\begin {equation}
\frac {\d x_1} {a_1} = \frac {\d x_2} {a_2} = \ldots = \frac {\d
x_n} {a_n}.\label {1.2}
\end {equation}

Provided sufficient information about the system \re {1.1}/\re
{1.2} is known, the multiplier is given by
\begin {equation}
\frac{\partial (f,
\omega_1,\omega_2,\ldots,\omega_{n-1})}{\partial(x_1,x_2,\ldots,x_n)}=MAf,
\label {1.4}
\end {equation}
where
\begin {equation}
\frac{\partial (f,
\omega_1,\omega_2,\ldots,\omega_{n-1})}{\partial(x_1,x_2,\ldots,x_n)}=
\mbox {\rm det}\left [\begin {array} {lcl}
\pa {f} {x_1} &\ldots &\pa {f} {x_n}\\
\pa {\omega_1} {x_1} & &\pa {\omega_1} {x_n}\\
\,\vdots & &\,\vdots\\
\pa {\omega_{n- 1}} {x_1} &\ldots &\pa {\omega_{n- 1}} {x_n}
\end {array}\right] \neq .  0\label {1.5}
\end {equation}
and $\omega_1 $, $\ldots $, $\omega_{n- 1} $ are $n- 1 $ solutions
of \re {1.1} or, equivalently, first integrals of \re {1.2}. As a
consequence, one can prove that each multiplier is a solution of
the partial differential equation
\begin {equation}
\sum_{i = 1} ^n\pa{\left (M a_i\right)} {x_i} = 0. \label {1.3}
\end {equation}

A different combination of the integrals can produce a different
multiplier, $M'$. The ratio $M'/M $ is a solution of \re {1.1} or
a first integral of \re {1.2}, which may be trivial as in the
application of the Poisson-Jacobi Theorem \cite{Poisson 09},
\cite{Jacobi40} to the determination of additional first
integrals. A  scholarly essay on the history of the Poisson-Jacobi
theorem which Jacobi considered \cite{Jacobi40}
\begin{verse} la plus profonde d\'ecouverte de M.
Poisson\footnote{`the most profound discovery by Mr Poisson'; see
also the letter of Jacobi \cite{Jacobi nn}}
 \end{verse} and the pervasive influence of Jacobi's work upon
Lie can be found in \cite{Hawkins}.

In its original form the method of Jacobi's last multiplier
required almost complete knowledge of the system under
consideration.  Since the existence of a first integral is
consequent upon the existence of symmetry, one is not surprised
that Lie \cite {Lie 12} [pp 333-347] provided a symmetric route to
the determination of Jacobi's last multiplier.  A more transparent
treatment is given by Bianchi \cite {Bianchi 18} [pp 456-464].

Suppose that we know $n- 1 $ symmetries of \re {1.1}/\re {1.2}
\begin {equation}
X_i = \xi_{ij}\partial_{x_j},\quad i = 1,n- 1. \label {1.6}
\end {equation}
Then Jacobi's last multiplier is also given by
\begin{equation}
M=\frac{1}{\Delta}
\end{equation}
 in the case that
$\Delta \neq 0 $,
 where now
\begin {equation}
\Delta = \mbox {\rm det}\left[
\begin {array}{llcl}
a_1, & a_2, &\ldots, & a_n\\
\xi_{11}, & & &\xi_{1,n}\\
\,\vdots & & &\,\vdots\\
\xi_{n- 1,1}, &\xi_{n- 1,2},&\ldots, &\xi_{n- 1,n}
\end {array}\right]. \label {1.7}
\end {equation}

Jacobi's last multiplier provides an incestuous interrelationship
between symmetries, first integrals and integrating factors for
well-endowed systems.  The practicality of this interrelationship
was somewhat diminished in the past due to the effort required to
evaluate the determinants of matrices of even moderate size.  For
example a Newtonian problem in three dimensions would require the
evaluation of the determinant of a $6\times 6 $ matrix ($7\times 7
$ if one considers time dependence, \ie $a_1=1$). This possibly
explains the omission of discussion of the method by postclassical
authors such as Cohen \cite{Cohen 11}, Dickson \cite{Dickson 24}
and Eisenhart \cite {Eisenhart 36}. The ready availability of
computer algebra systems has rendered this method an attractive
alternative, say, for the determination of first integrals given
symmetries.  To take a trivial example the `free particle' has the
Newtonian equation
\begin {equation}
\ddot {y} = 0 \quad \Leftrightarrow \quad \left \{\begin {array}
{l}
\dot {u}_1 = u_2\\
\dot {u}_2 = 0
\end {array} \right .\label {1.8}
\end {equation}
with the Lie point symmetries
\begin {equation}
\begin {array} {ll}
\Gamma_1 = \partial_{u_1} &\Gamma_5 = t\partial_t-u_2\partial_{u_2}\\
\Gamma_2 = t\partial_{u_1} +\partial_{u_2}
&\Gamma_6 = t ^ 2\partial_t+tu_1\partial_{u_1} +\left (u_1-tu_2\right)\partial_{u_2}\\
\Gamma_3 = u_1\partial_{u_1} +u_2\partial_{u_2}\quad\quad
&\Gamma_7 = u_1\partial_t-u_2^2\partial_{u_2}\\
\Gamma_4 = \partial_t &\Gamma_8 =tu_1\partial_t+u_1 ^
2\partial_{u_1}
 +\left (u_1-tu_2\right)u_2\partial_{u_2}.
\end {array}\label{1.9}
\end {equation}
We find, for example, the determinants
\begin {displaymath}
\begin {array} {lcl}
\Delta_1  & = &  \left |\begin {array} {lll}
1 & u_2 & 0\\
0 & t & 1\\
0 & u_1 & u_2
\end {array}\right | = tu_2-u_1\nonumber \\\nonumber\\
\Delta_2 & = &\left | \begin {array} {lll}
1 & u_2 & 0\\
0 &  u_1 & u_2\\
t ^ 2 & tu_1 & u_1-tu_2
\end {array}\right | = \left (u_1-tu_2\right) ^ 2\nonumber\\\nonumber\\
\Delta_3 & = &\left |\begin {array} {lll}
1 & u_2 & 0\\
1 & 0 & 0\\
tu_1 & u_1 ^ 2 & \left (u_1-tu_2\right)u_2
\end {array}\right | = -\left.
(u_1-tu_2\right)u_2^2,
\end {array} \label {1.10}
\end {displaymath}
using $\Gamma_2 $ and $\Gamma_3 $; $\Gamma_3 $ and $\Gamma_6 $;
and $\Gamma_4 $ and $\Gamma_8 $ respectively, and the integrals
\begin {gather}
I_1 = \frac {\Delta_2} {\Delta_1} = tu_2-u_1\nonumber\\
I_2 =\frac {\Delta_3} {\Delta_1} = u_2^2 \label{1.11}
\end {gather}
as expected. (Note that there are twenty-eight possible
determinants to be calculated.  Some of these are zero.)

\section {Complete Symmetry Groups}

In 1994 Krause \cite {Krause 94} introduced a new concept of the
complete symmetry group of a system by defining it as the group
represented by the set of symmetries required to specify the
system completely.  There is not necessarily any relationship
between the symmetries required to specify completely a system and
its point symmetries.  Thus Andriopoulos \etal \cite {Andriopoulos
01} reported the complete symmetry group of the `free particle' to
be $A_{3,3} $ ($D\oplus_s T_2 $), the semidirect sum of dilations
and translations in the plane, with the symmetries being three of
the usual eight Lie point symmetries of (\ref {1.8} a), and that
of the Ermakov-Pinney equation \cite {Ermakov 80, Pinney 50} to be
$A_{3,8} $ ($sl (2,R) $) with the symmetries being the three Lie
point symmetries of that equation.  On the other hand Krause \cite
{Krause 94} reported that an additional three nonlocal symmetries
are necessary to specify the Kepler Problem completely since the
five Lie point symmetries of the three-dimensional Kepler Problem
are insufficient to the purpose. This contains the implication
that eight Lie symmetries are necessary to specify completely the
Kepler Problem.  However, a more careful analysis of sufficiency
by Nucci \etal\ \cite {Nucci 02 a}, based on the method of
reduction of order proposed by Nucci in 1996 \cite {Nucci96b} and
her interactive code for the determination of Lie symmetries \cite
{Nucci 90, Nucci 96 a}, has revealed that the equation for the
three-dimensional Kepler Problem is completely specified by six
Lie symmetries with the algebra $A_1\oplus\left
\{A_1\oplus_s\{2A_1\oplus 2 A_1\right\}\} $ ($\Leftrightarrow
A_1\oplus\left\{D\oplus_s\{T_2\oplus T_2\right\}\} $), where the
subalgebra $D\oplus_s T_2 $ is that associated with the first
integrals of the one-dimensional simple harmonic oscillator \cite
{Mahomed 88} (equally any second order ordinary differential
equation with the algebra $sl (3,R) $) to which the Kepler Problem
reduces naturally \cite {Nucci 01} under the method of reduction
of order.

One assumes that, when he devised his method of the last
multiplier, the original intention of Jacobi was to determine
integrating factors and that the adaptation from integral to
symmetries by Lie was of like intention.  However, \re {1.7} in
combination with \re {1.3} suggests the possibility to determine
symmetries provided the multiplier is known.  The general solution
of \re {1.3} is equivalent to the solution of \re {1.2} unless one
has the opportunity to perceive a particular solution without real
effort.  A particular case in point is when the functions $a_i (x)
$ are independent of $x_i $ for then \re {1.3} has the solution
that $M $ is a constant, taken to be chosen as a convenient value,
which in this instance is not a `trivial' solution.  Since one now
has an $M $, one can attempt to determine a further symmetry by
solving \re {1.7} with one row of the matrix the coefficient
functions of the unknown symmetry.  One may infer that Lie and
Bianchi had in mind point and contact symmetries in their
treatments of Jacobi's last multiplier from the basis of
symmetries rather than the original approach through first
integrals used by Jacobi.  However, as is common with many of the
theoretical properties and applications of symmetries, there is no
statement of the variable dependence of the coefficient functions
in the method of Jacobi's last multiplier required for the method
to hold.  Consequently the considerations above apply equally to
determination of nonlocal symmetries, in particular nonlocal
symmetries of the type used by Krause, in which the nonlocality is
found in the coefficient function of the independent variable, for
autonomous systems.  For such systems one of the known symmetries
is $\partial_t $ which is represented in the matrix of \re {1.7}
by the row $(1,0,\ldots, 0) $.  In the Laplace expansion of the
determinant the only possible nonzero terms must contain the first
element of this row.  If the unknown symmetry is
\begin {equation}
\Gamma_n = V\partial_t+ G_i\partial_{u_i}, \label {2.1}
\end {equation}
$V $ does not appear in the expression for $\Delta $, only the
$G_i $. These may be selected at will to satisfy the requirement
that $\Delta = 1 $, or other suitable constant, and for each
selection a $V $ be computed through the invariance of the system
of first order ordinary differential equations under the action of
$\Gamma_n ^ {[1]} $, the first extension of $\Gamma_n $.  In
principle this would permit $n $ symmetries to be determined.
However, that presumes the independence of the $G_i $, $i = 1,n $.
This need not be the case if the $u_i $ come from the reduction of
an $n $th order scalar ordinary differential equation and the
imposition of point symmetries is made at the level of the $n $th
order equation.
\section {The complete symmetry group of the Euler-Poinsot system}

As an illustration of the ideas contained in \S\S 1 and 2 we
consider the simplest case of the motion of a rigid body which is
the system governed by the Euler-Poinsot equations \cite {Euler,
Poinsot}
\begin {gather}
\dot {\omega}_1 = \frac {B-C} {A}\omega_2\omega_3 = W_1\nonumber\\
\dot {\omega}_2 = \frac {C- A} {B}\omega_3\omega_1 = W_2 \label {3.1}\\
\dot {\omega}_3 = \frac {A -B} {C}\omega_1\omega_2 = W_3\nonumber
\end {gather}
in which $\mbox{\boldmath{$\omega$}}: =
(\omega_1,\omega_2,\omega_3) ^T $ is the angular velocity and $A
$, $B $ and $C $ are the principal moments of inertia.  It is a
commonplace that the system \re {3.1} possesses respectively the
two first integrals and Lie point symmetries
\begin {equation}
\begin {array} {ll}
E = \half\left (A\omega_1 ^ 2+ B\omega_2 ^ 2+ C\omega_3 ^
2\right)\quad\quad & L ^ 2 = A ^ 2\omega_1 ^ 2+ B ^ 2\omega_2 ^ 2+
C ^ 2\omega_3 ^ 2
\\
\Gamma_1 = \partial_t &\Gamma_2 =
-t\partial_t+\omega_i\partial_{\omega_i}.
\end {array}
\end {equation}
We note that Nucci's method of reduction of order \cite {Nucci96b}
looks for first integrals in which one variable is missing
\cite{MarcelNuc} and provides \cite{lorpoin}
\begin {gather}
I_1 = \omega_2 ^ 2B (A -B) - \omega_3 ^ 2C (C- A)\nonumber\\
I_2 = \omega_3 ^ 2C (B-C) - \omega_1 ^ 2 A (A -B) \label{3.4}\\
I_3 = \omega_1 ^ 2 A (C- A) - \omega_2 ^ 2B (B-C)\nonumber
\end {gather}
as three conserved quantities, which are obviously not independent
and can be constructed from the two independent integrals, $E $
and $L ^ 2$, by the respective elimination of $\omega_1 $,
$\omega_2 $ and $\omega_3 $ from them.

We note that $\omega_i $ is absent from the right hand side of
$\dot {\omega}_i $ in \re {3.1} and so one solution of \re {1.3}
for Jacobi's last multiplier is a constant.  The condition that
$\Gamma_n $  \re {2.1} be a third symmetry of the Euler-Poinsot
system \re {3.1} is
\begin {equation}
\Delta = \mbox {\rm Det}\left [
\begin {array} {rlll}
1 &W_1 &W_2 &W_3 \\
1 & 0 & 0 & 0\\
-t &\omega_1 &\omega_2  &\omega_3\\
V &G_1 &G_2 &G_3
\end {array}\right]
= a, \label {3.5}
\end {equation}
where $a $ is the constant to which we may assign some convenient
value.  We obtain
\begin {equation}
A\omega_1 I_1G_1+ B\omega_2 I_2G_2+ C\omega_3 I_3G_3 = a ABC.
\label {3.6}
\end {equation}

We take
\begin {equation}
G_1 = \frac {aBC} {\omega_1 I_1},\quad G_2 = 0,\quad G_3 =
0\quad\quad\mbox {\it et cyc}\label {3.7}
\end {equation}
and set $a = I_1/BC $ {\it et cyc} to obtain the three sets of
coefficient functions for the dependent variables
\begin {equation}
\left (\frac {1} {\omega_1}, 0,0\right);\quad\left (0,\frac {1}
{\omega_2}, 0\right);\quad\left (0,0,\frac {1}
{\omega_3}\right).\label {3.8}
\end {equation}
We note that the sets in \re {3.8} provide a basis and that other
combinations could be taken.
 We keep the forms \re {3.8} simply for their present simplicity and find their subsequent
 utility.

It remains to determine $V $.  Consider the first set in \re
{3.8}. The corresponding symmetry is written as
\begin {gather}
\Gamma_3 = V_3\partial_t+\frac {1} {\omega_1}\partial_{\omega_1} \label {3.9}\\
\Gamma_3 ^ {[1]} = \Gamma_3-\dot {\omega}_1\left (\frac {1}
{\omega_1 ^ 2} +\dot {V}_3\right)\partial_{\dot {\omega}_1} -\dot
{\omega}_2\dot {V}_3
\partial_{\dot {\omega}_2} -\dot {\omega}_3\dot {V}_3\partial_{\dot {\omega}_3} \label {3.10}
\end {gather}
and the action of $\Gamma_3 ^ {[1]} $ on the Euler-Poinsot system,
\re {3.1}, is
\begin {gather}
-\dot {\omega}_1\left (\frac {1} {\omega_1 ^ 2} +\dot {V}_3\right) = 0\nonumber\\
\frac {\dot {\omega}_2} {\omega_1 ^ 2} = \frac {C- A} {B}\frac {\omega_3} {\omega_1}  \label{3.11}\\
\frac {\dot {\omega}_3} {\omega_1 ^ 2} = \frac {A -B} {C}\frac
{\omega_2} {\omega_1}\nonumber
\end {gather}
in which (\ref {3.11}b,c) are consistent with \re {3.1} and (\ref
{3.11}a) gives
\begin {equation}
V_3 = -\int\frac {\dt} {\omega_1 ^ 2}. \label{3.122}
\end {equation}
A similar calculation applies to second and third of \re {3.8}.

We obtain the three nonlocal symmetries
\begin {gather}
\Gamma_3 = -\left (\int\frac {\dt} {\omega_1 ^ 2}\right)\partial_t+\frac {1} {\omega_1}\partial_{\omega_1} \nonumber\\
\Gamma_4 = -\left (\int\frac {\dt} {\omega_2 ^ 2}\right)\partial_t+\frac {1} {\omega_2}\partial_{\omega_2} \label {3.12}\\
\Gamma_5 = -\left (\int\frac {\dt} {\omega_3 ^
2}\right)\partial_t+\frac {1} {\omega_3}\partial_{\omega_3}
\nonumber
\end {gather}
for the Euler-Poinsot system \re {3.1}.

Since the system \re {3.1} is of the first order and autonomous,
it is in a suitable form for reduction of order.  We set $y =
\omega_3 $ as the new independent variable.  This is an arbitrary
choice.  Equally $\omega_1 $ or $\omega_2 $ could be chosen as the
independent variable.  The results do not differ.  The reduced
system is
\begin {gather}
\dm{\omega_1} {y} = \frac {(B-C)C} {A (A -B)}\frac {y} {\omega_1}\nonumber\\
\dm{\omega_2} {y} = \frac {(C- A)C} {B (A -B)}\frac {y} {\omega_2}
\label {3.13}
\end {gather}
and inherits the symmetries
\begin {gather}
\tilde {\Gamma}_2 = \omega_1\partial_{\omega_1} +\omega_2\partial_{\omega_2} + y\partial_y\nonumber\\
\tilde {\Gamma}_3 = \frac {1} {\omega_1}\partial_{\omega_1}\nonumber\\
\tilde {\Gamma}_4 = \frac {1} {\omega_2}\partial_{\omega_2}\nonumber\\
\tilde {\Gamma}_5 = \frac {1} {y}\partial_{y} \label {3.14}
\end {gather}
with the Lie Brackets
\begin {gather}
\left [\tilde {\Gamma}_2,\tilde {\Gamma}_3\right] = -2\tilde {\Gamma}_3,\quad \left [\tilde {\Gamma}_3,\tilde {\Gamma}_4\right] = 0,\quad \left [\tilde {\Gamma}_4,\tilde {\Gamma}_5\right] = 0\nonumber\\
\left [\tilde {\Gamma}_2,\tilde {\Gamma}_4\right] = -2\tilde {\Gamma}_4,\quad \left [\tilde {\Gamma}_3,\tilde {\Gamma}_5\right] = 0, \label {3.15}\\
\left [\tilde {\Gamma}_2,\tilde {\Gamma}_5\right] = -2\tilde
{\Gamma}_5.\nonumber
\end {gather}

Consider a general two-dimensional system
\begin {gather}
\dm {\omega_1} {y} = f_1 (\omega_1,\omega_2,y)\nonumber\\
\dm {\omega_2} {y} = f_2 (\omega_1,\omega_2,y),\label {4.1}
\end {gather}
of which the system \re {3.13} is a specific instance.  We
determine which of the four symmetries $\tilde
{\Gamma}_2,\ldots,\tilde {\Gamma}_5 $ are necessary to specify \re
{3.13} given \re {4.1}.  (This does beg the question of the
appropriateness of these four symmetries, but this will eventually
become apparent.)

The actions of
\begin {gather}
\tilde {\Gamma}_2 ^ {[1]} = \omega_1\partial_{\omega_1} +\omega_2\partial_{\omega_2} +y\partial_y\nonumber\\
\tilde {\Gamma}_3 ^ {[1]} = \frac {1}
{\omega_1}\partial_{\omega_1}
  -\frac {\omega_1'} {\omega_1 ^ 2}\partial_{\omega_1'}\label {4.2}\\
\tilde {\Gamma}_4 ^ {[1]} = \frac {1}
{\omega_2}\partial_{\omega_2} -\frac {\omega_2'} {\omega_2 ^
2}\partial_{\omega_2'}\nonumber
\end {gather}
on (\ref {4.1} a) give
\begin {gather}
0 = y\pa {f_1} {y} +\omega_1\pa {f_1} {\omega_1} +\omega_2\pa {f_1} {\omega_2}\nonumber\\
-\frac {f_1} {\omega_1 ^ 2} = \frac {1}{\omega_1}\pa {f_1} {\omega_1}\label {4.3}\\
0 = \frac {1} {\omega_2}\pa {f_1} {\omega_2}\nonumber
\end {gather}
from which it is evident that
\begin {equation}
f_1 (\omega_1,\omega_2,y) = K\frac {y} {\omega_1}\label {4.4}
\end {equation}
and hence (\ref {3.13} a) is recovered.

The same symmetries acting on (\ref {4.1}b) lead to (\ref {3.13}b)
and so the three symmetries, $\tilde {\Gamma}_2 $, $\tilde
{\Gamma}_3 $ and $\tilde {\Gamma}_4 $ are a representation of the
complete symmetry group of the system \re {3.13}.  By means of a
similar calculation we see that the triplets of $\tilde {\Gamma}_2
$, $\tilde {\Gamma}_3 $ and $\tilde {\Gamma}_5 $ and of $\tilde
{\Gamma}_2 $, $\tilde {\Gamma}_4 $ and $\tilde {\Gamma}_5 $ are
also representations of the complete symmetry group of \re {3.13}.

The listing of Lie Brackets in \re {3.15} shows that the Lie
algebra in each of the three cases is $A_1\oplus_s 2 A_1 $ or
$D\oplus_s T_2 $, a representation of the pseudo-Euclidean group
$E (1, 1) $ of dilations and translations in the plane.

This is not be end of the story.  Consider about the actions of
$\tilde {\Gamma}_3 ^ {[1]} $, $\tilde {\Gamma}_4 ^ {[1]} $ and
$\tilde {\Gamma}_5 ^ {[1]} $ on (\ref {4.1} a).  We obtain three
constraints on $f_1 $, \viz
\begin {gather}
-\frac {f_1} {\omega_1 ^ 2} = \frac {1} {\omega_1}\pa {f_1} {\omega_1}\nonumber\\
0 = \frac {1} {\omega_2}\pa {f_1} {\omega_2}\label {4.5}\\
\frac {f_1} {y ^ 2} = \frac {1} {y}\pa {f_1} {y}\nonumber
\end {gather}
when (\ref {4.1} a) is taken into account.  It is obvious that
(\ref {3.13} a) is recovered. Similarly (\ref {4.1}b) reduces to
(\ref {3.13}b).

Consequently the three symmetries $\tilde {\Gamma}_3 $, $\tilde
{\Gamma}_4 $ and $\tilde {\Gamma}_5 $ provide a representation of
the complete symmetry group of \re {3.13}.  In this case the
algebra of the symmetries is the abelian $3 A_1 $\footnote {At
first glance the existence of $3 A_1 $ for a second order system
may seem to be at odds with the general result that a second order
ordinary differential equation -- a second order system -- cannot
admit $3 A_1 $.  However, \re {3.13} cannot be written as a scalar
second order ordinary differential equation. This emphasises the
general point that a higher-order scalar equation may be reduced
to a system of first order ordinary differential equations, but
the reverse process is not always possible.} and not the
$A_1\oplus_s 2 A_1 $ of the previous algebras.

The system \re {3.13} is {\it not} the system \re {3.1} and one
cannot {\it expect} that the complete symmetry group of \re {3.1}
would be that of \re {3.13} although from the point of view of
differential equations both are equally described in terms of a
three-dimensional phase space.  We consider the two triplets, \viz
$\Gamma_3 $, $\Gamma_4 $ and $\Gamma_5 $ and $\Gamma_2 $,
$\Gamma_3 $ and $\Gamma_4 $ (equivalently $\Gamma_2 $, $\Gamma_3 $
and $\Gamma_5 $ and $\Gamma_2 $, $\Gamma_4 $ and $\Gamma_5 $ as we
saw above) and their respective actions on the general system.
\begin {gather}
\dot {\omega}_1 = f_1 (t,\omega_1,\omega_2,\omega_3)  \nonumber\\
\dot {\omega}_2 = f_2 (t,\omega_1,\omega_2,\omega_3)  \label {4.6}\\
\dot {\omega}_3 = f_3 (t,\omega_1,\omega_2,\omega_3)  \nonumber
\end {gather}
to which class the Euler-Poinsot system \re {3.1} belongs.

The actions of the first extensions of $\Gamma_3 $, $\Gamma_4 $
and $\Gamma_5 $, \viz
\begin {gather}
\Gamma_3 ^ {[1]} =\left (-\int\frac {\dt} {\omega_1 ^ 2}\right)\partial_t+\frac {1} {\omega_1}\partial_{\omega_1} +\frac {\dot {\omega}_2} {\omega_1 ^ 2}\partial_{\dot {\omega}_2} +\frac {\dot {\omega}_3} {\omega_1 ^ 2}\partial_{\dot {\omega}_3}  \nonumber  \\
\Gamma_4 ^ {[1]} =\left (-\int\frac {\dt} {\omega_2 ^ 2}\right)\partial_t+\frac {1} {\omega_2}\partial_{\omega_2} +\frac {\dot {\omega}_1} {\omega_2 ^ 2}\partial_{\dot {\omega}_1} +\frac {\dot {\omega}_3} {\omega_2 ^ 2}\partial_{\dot {\omega}_3}  \label {4.7}  \\
\Gamma_5 ^ {[1]} =\left (-\int\frac {\dt} {\omega_3 ^
2}\right)\partial_t+\frac {1} {\omega_3}\partial_{\omega_3} +\frac
{\dot {\omega}_1} {\omega_3 ^ 2}\partial_{\dot {\omega}_1} +\frac
{\dot {\omega}_2} {\omega_3 ^ 2}\partial_{\dot {\omega}_2},
\nonumber
\end {gather}
on (\ref {4.6} a) are
\begin {gather}
0 = -T_1\pa {f_1} {t} +\frac {1} {\omega_1}\pa {f_1} {\omega_1}\nonumber\\
\frac {f_1} {\omega_2 ^ 2} = - T_2\pa {f_1} {t} +\frac {1} {\omega_2}\pa {f_1} {\omega_2}\label {4.8}\\
\frac {f_1} {\omega_3 ^ 2} = - T_3\pa {f_1} {t} +\frac {1}
{\omega_3}\pa {f_1} {\omega_3}\nonumber
\end {gather}
in which we have written $T_i =\int\dt/\omega_i ^ 2 $.  The system
\re {4.8} does not contain sufficient information to reduce (\ref
{4.6} a) to the first of system \re {3.1}.  The same applies for
(\ref {4.6}b) and (\ref {4.6}c).  The abelian group $3 A_1 $
represented by $\Gamma_3 $, $\Gamma_4 $ and $\Gamma_5 $ is not the
complete symmetry group of \re {3.1}.

The first extensions of $\Gamma_2 $ is
\begin {equation}
\Gamma_2 ^ {[1]} = -t\partial_t+\omega_1\partial_{\omega_1}
+\omega_2\partial_{\omega_2} +\omega_3\partial _{\omega_3} + 2\dot
{\omega_1}\partial_{\dot {\omega}_1} + 2\dot
{\omega}_2\partial_{\dot {\omega}_3}.\label {4.9}
\end {equation}
The actions of this, $\Gamma_3 ^ {[1]} $ and $\Gamma_4 ^ {[1]} $
on (\ref {4.6}a) are
\begin {gather}
2f_1 = -t\pa {f_1} {t} +\omega_1\pa {f_1} {\omega_1} +\omega_2\pa {f_1} {\omega_2} +\omega_3\pa {f_1} {\omega_3}\nonumber\\
0 = - T_1\pa {f_1} {t} +\frac {1} {\omega_1}\pa {f_1} {\omega_1}\label {4.10}\\
\frac {f_1} {\omega_2 ^ 2} = - T_2\pa {f_1} {t} +\frac
{1}{\omega_2}\pa {f_1} {\omega_2}.\nonumber
\end {gather}

From the first of \re {4.10}
\begin {equation}
f_1 = t ^ {- 2}F_1 (u,v,w),\label {4.11}
\end {equation}
where $u $, $v $ and $w $ are the three characteristics
independent of $f_1 $, \viz $t\omega_1 $, $t\omega_2 $ and
$t\omega_3 $.  The substitution of \re {4.11} into (\ref {4.10}b)
and (\ref {4.10}c) gives, respectively,
\begin {gather}
\frac {1} {u}\pa {F_1} {u} = \frac {T_1} {t ^ 3}\left [- 2F_1+u\pa
{F_1} {u} +v\pa {F_1} {v}
 +w\pa {F_1} {w}\right]\nonumber\\
\frac {1} {v}\pa {F_1} {v} -\frac {F_1} {v ^ 2} = \frac {T_2} {t ^
3}\left [- 2F_1+u\pa {F_1} {u} +v\pa {F_1} {v} +w\pa {F_1}
{w}\right]\label {4.12}
\end {gather}
and now the situation is entirely different since the $t $
dependence outside of the characteristics is isolated in the
coefficients of the terms within crochets in both (\ref {4.12} a)
and (\ref {4.12}b).  Consequently we have the three terms
separately zero, \ie
\begin {gather}
\frac {1} {u}\pa {F_1} {u} = 0\nonumber\\
\frac {1} {v}\pa {F_1} {v} -\frac {F_1} {v ^ 2}= 0\label {4.13}\\
u\pa {F_1} {u} +v\pa {F_1} {v} +w\pa {F_1} {w} = 2F_1.\nonumber
\end {gather}
We recover (\ref {3.1}a).  Like calculations recover (\ref {3.1}b)
and (\ref {3.1}c).

The complete symmetry group of the Euler-Poinsot system, \re
{3.1}, is $ E (1,1) $ ($\Leftrightarrow D\otimes_s T_2 $).  There
are three equivalent representations.

\section {Discussion}

In this paper we have brought together several disparate ideas to
arrive at something of a question mark.  The symmetry-based
version of Jacobi's last multiplier has been turned on its head,
as it were, to provide a means to calculate nonlocal symmetries,
in this instance, for the Euler-Poinsot system given the last
multiplier.  With these nonlocal symmetries we can identify the
set of symmetries which specify completely the Euler-Poinsot
system from all possible classes of systems of three first order
equations.  The number of symmetries required for this
specification is three.

This is surprising in terms of general expectations.  The general
first order equation for a system with one independent and $n $
dependent variables is
\begin {equation}
\dot {\omega}_i = f_i (t,\omega_1,\ldots,\omega_n).\label {4.19}
\end {equation}

With the application of each symmetry the number of variables in
$f_i $ is reduced by one so that, after the application of $n $
symmetries, there remains a general function of one
characteristic.  The application of a further symmetry specifies
that general function.  Thus we would expect that the complete
symmetry group would have a representation in terms of an algebra
of $n+ 1 $ elements.  In fact one would regard an algebra of fewer
than $n+ 1 $ elements as being quite unusual since it would imply
an unexpected degree of connectedness amongst the coefficient
functions of the different variables in the symmetries.  (Such an
example is found most dramatically in the case of the Kepler
Problem \cite {Nucci 02 a}.)

It is already known \cite {Andriopoulos 01} that a given system
may possess more than one representation of its complete symmetry
group.  Indeed this is quite standard for any equation of the
second order possessing eight symmetries.

In the case of the Euler-Poinsot system we have found a rather
intriguing result. In the reduced two-dimensional system, \re
{3.13}, we found that they were two inequivalent representations
of the complete symmetry group, \viz $A_1\oplus_s 2 A_1 $ and $3
A_1 $.  This lack of uniqueness does not persist when one returns
to the three-dimensional system, properly known as the
Euler-Poinsot system, for then at the latter symmetry group, \viz
$3A_1$, falls away as a representation of the complete symmetry
group.  We do find that in conjunction with $\Gamma_1 $ the three
symmetries of $3 A_1 $ do specify \re {3.1}, but the
four-dimensional algebra is not a candidate as a representation of
the complete symmetry group since its dimensionality is not
minimal.

It would be interesting to find other examples of systems
exhibiting similar properties\footnote {For a recent instance of
which see the paper
 by Andriopoulos \etal \cite {Andriopoulos 02 a} in the present volume.}.
Certainly the present result does place something of a question
mark against the interpretation of the concept of a complete
symmetry group as being {\it the} group of the symmetries which
completely specify the equation although it does this without
detracting from the inherent interest of the concept of a complete
symmetry group.  In fact one must seriously consider the
identification of the characteristic system for a given problem,
in this case whether it is the three-dimensional system \re {3.1},
which is the standard Euler-Poinsot system, or the reduced
two-dimensional system \re {3.13}.  Curiously enough a similar
question has arisen in the case of the Painlev\'e Property for
systems of first-order ordinary differential equations \cite
{Leach 01}.

\subsection*{Acknowledgements}

PGLL thanks the Dipartimento di Matematica e Informatica,
Universit\`a di Perugia, and Dr MC Nucci for their kind
hospitality while this work was undertaken and the National
Research Foundation of South Africa and the University of Natal
for their continuing support.

\label{lastpage}

\end{document}